\documentclass[12pt]{article}
\usepackage{amssymb}

\newcommand{\be}{\begin{equation}}
\newcommand{\ee}{\end{equation}}
\newcommand{\bea}{\begin{eqnarray}}
\newcommand{\eea}{\end{eqnarray}}

\begin{document}

\bigskip 
\begin{titlepage}

\begin{flushright}
UUITP-01/03\\ 
hep-th/0301182
\end{flushright}

\vspace{1cm}

\begin{center}
{\Large\bf Holography, inflation, and quantum fluctuations in the early universe\\}

\end{center}
\vspace{3mm}

\begin{center}

{\large
Ulf H.\ Danielsson} \\

\vspace{5mm}

Institutionen f\"or Teoretisk Fysik, Box 803, SE-751 08
Uppsala, Sweden

\vspace{5mm}

{\tt
ulf@teorfys.uu.se \\
}

\end{center}
\vspace{5mm}

\begin{center}
{\large \bf Abstract}
\end{center}
\noindent
In this paper the relevance of holographic entropy bounds in the context of inflation is 
investigated.  We distinguish between entropy on 
large and small scales and confront the entropy
of quantum fluctuations in an inflating cosmology with the appropriate entropy bounds. 
In conclusion we do not find any constraints on inflation from holography, but some suggestions for
future studies are given.

%unstable 

\vfill
\begin{flushleft}
January 2003
\end{flushleft}
\end{titlepage}
\newpage

%%%% INTRODUCTION

\section{Introduction}

\bigskip

One of the most intriguing challenges in modern physics is to find
observable consequences of quantum gravity. Recently the attention has
focused towards cosmology, where in particular inflation might serve as a
useful testing ground for new ideas. One proposal is that the magnifying
effect of inflation might allow physics beyond the Planck scale to show up
in the CMBR-spectrum. For reviews see \cite{Brandenberger:2002sr}\cite
{kinney2003}. Another approach, based on the successful introduction of
holography in string theory, aims at finding meaningful constraints on
inflation through holography. Some years ago there were several attempts in
this direction, but the general conclusion, \cite{Easther:1999gk}\cite
{Veneziano:1999ts}\cite{Kaloper:1999tt}, was that a correct implementation
of holography did not give anything more than what follows from the
generalized second law of thermodynamics \cite{Bekenstein:ax}.

More recently there has been several interesting works discussing various
aspects of holography and cosmology, \cite{Banks:2000fe}\cite{Banks:2001yp} 
\cite{Hellerman:2001yi}\cite{Fischler:2001yj}\cite{Banks:2001px} \cite
{Witten:2001kn}\cite{Dyson:2002nt}\cite{Dyson:2002pf}\cite{Myung:2003hb},
where many conceptual problems of physics in de Sitter space are addressed.
Furthermore, in \cite{Hogan:2002xs} and \cite{Albrecht:2002xs}, it has been
argued that there are, in fact, situations where holography actually do
imply non trivial constraints on inflation and that the effects might even
be detectable. Other groups, like \cite{Frolov:2002va}, do not find any such
constraints on inflationary physics from holography and are therefore more
in line with the earlier investigations.

In the present paper we will find evidence in favor of the conservative
point of view. We will discuss the various entropy bounds during inflation
and explain why they do not seem to yield any new constraints. The outline
is as follows. In section two we review the Hubble and D-bounds for the
entropy in an expanding universe. In section three we investigate more
carefully \textit{where} the entropy is from the point of view of different
observers. In particular we pay attention to the scale on which the entropy
can be found. We also discuss the entropy of the quantum fluctuations
responsible for the fluctuations in the CMBR. In section four we investigate
in more detail the entropy bounds in a slowly rolling universe without
finding any constraints from holography. We end, in section five, with some
speculations on how to find other ways of applying holography to inflation.

\section{The Hubble and D-bound}

\bigskip

The entropy bound that will serve as a starting point for our discussion is
the \textit{Bekenstein bound} in asymptotically flat space, \cite
{Bekenstein:jp}, which states that 
\begin{equation}
S\leq S_{B}=2\pi ER,
\end{equation}
where $E$ is the energy contained in a volume with radius $R$. There are
several arguments in support of the bound when gravity is weak \cite
{Bekenstein:vm}\cite{Schiffer:et}, and it is widely believed to hold true
for all reasonable physical systems. Furthermore, in the case of a black
hole where $R=2El_{p}^{2}$, we have an entropy given by 
\begin{equation}
S_{BH}=\frac{A}{4l_{p}^{2}}=\frac{\pi R^{2}}{l_{p}^{2}},
\end{equation}
which exactly saturates the Bekenstein bound. We will consequently put $%
\hbar =c=1$, but explicitly write the Planck length, $\ l_{p}=\sqrt{\frac{%
G\hbar }{c^{3}}}$, to keep track of effects due to gravity.

Beginning with \cite{Fischler:1998st}, there have been many attempts to
apply similar entropy bounds to cosmology and in particular to inflation.
(See for instance \cite{KalyanaRama:1998pk}.) The idea has been to choose an
appropriate volume and argue that the entropy contained within the volume
must be limited by the area. An obvious problem in a cosmological setting
is, however, that for a constant energy density a bound of this type always
will be violated if the radius $R$ of the volume is chosen to be big enough.
However, as was explained in \cite{Easther:1999gk}\cite{Veneziano:1999ts}
\cite{Bak:1999hd}, it is not reasonable to discuss radii which are larger
than the Hubble radius in the expanding universe. See also \cite
{Brustein:1999ua}\cite{Brustein:1999ay}\cite{Brustein:1999md}\cite
{Brustein:2000fw}. This, then, suggests that the maximum entropy in a volume
of radius $R>r$, where $r$ is the Hubble radius, is obtained by filling the
volume with as many Hubble volumes as one can fit -- all with a maximum
entropy of $\frac{\pi r^{2}}{l_{p}^{2}}$. This gives rise to the \textit{%
Hubble bound}, which states that 
\begin{equation}
S<S_{H}\sim \frac{R^{3}}{r^{3}}\frac{r^{2}}{l_{p}^{2}}=\frac{R^{3}}{%
rl_{p}^{2}}.
\end{equation}
As an illustrative example of how the Hubble bound works, we might consider
a radiation field with a temperature. We begin by considering a case where
the Hubble radius is not \ changing -- not even through slow roll. The
maximum entropy of the gas (attained in thermal equilibrium) is given by 
\[
S_{g}\sim \left( ER\right) ^{3/4}.
\]
This is clearly much smaller than the naive Bekenstein bound given by $%
S=2\pi ER$. But the important point in a cosmological setting, is that there
is nothing that prevents the energy $E$ to exceed the critical value $\frac{R%
}{2l_{p}^{2}}$, which in flat space would have corresponded to the formation
of a black hole. In other words, the entropy can easily exceed the value
given by $S_{BH}$, even if it can not exceed the Hubble bound $S_{H}$. In
fact, if the matter is in the form of a gas it can not even saturate the
bound. This follows since the entropy can not reach the Hubble bound value $%
S_{H}$ until the energy density is a factor $\left( \frac{r}{l_{p}}\right)
^{2/3}$ larger than the energy density $\frac{1}{r^{2}l_{p}^{2}}$ in the
cosmological constant. For saturation of the Hubble bound one would,
instead, need a collection of black holes.

The Hubble bound \ is a bound on the entropy that can be contained in a
volume much larger than the Hubble radius. It is, therefore, a bound that is
meaningful only from the point of view of a global observer, typically using
the standard, time dependent, FRW-coordinates. In order to be able to make
actual measurements of associated quantities one needs, therefore, inflation
to eventually stop and a non-accelerating phase to take over. The notion of
a cosmological horizon and the corresponding area does not play an important
role from this point of view since all entropy is present in matter,
possibly in the form of black holes.

If we, on the other hand, want to discuss things from the point of view of
what a local observer, that do not have time to wait for inflation to end,
can measure, we must be more careful. In this case one has a cosmological
horizon with an area that it is natural to give an entropic interpretation 
\cite{GibbHawk:1977}. Since the area of the horizon grows when matter is
passing out towards the horizon, from the point of view of the local
observer, it is natural to expect the horizon to encode information about
matter that, in its own reference frame, has passed to the \textit{outside}
of the cosmological horizon of the local observer. From the point of view of
the observer, the matter will never be seen to leave but rather become more
and more redshifted. The outside of the cosmological horizon should,
therefore, be compared with the inside of a black hole. It follows that the
horizon only indirectly provides bounds on entropy within the horizon as is
nicely exemplified through the \textit{D-bound} introduced in \cite
{Bousso:2000md}. The cosmological horizon area in a de Sitter space with
some extra matter is smaller than the horizon area in empty space. If the
matter passes out through the horizon, the increase in area can be used to
limit the entropy content in matter. This is the content of the D-bound
which turns out to coincide with the Bekenstein bound. The D-bound,
therefore has not, necessarily, that much to with de Sitter space or
cosmology. It is more a way to use de Sitter space to derive a constraint on
matter itself.

\bigskip

\section{Entropy in an expanding universe}

\bigskip

\subsection{Where is the entropy?}

\bigskip

In this section we will try to understand the nature and relations between
the various entropy bounds a little bit better. In particular we must find
out on what scales the entropy is stored. If we assume that all entropy is
stored on short scales smaller than the horizon scale $r$, we can consider
each of the horizon bubbles separately and use the Bekenstein bound (or
D-bound) on each and everyone of these volumes. We conclude from this that
the entropy, under the condition that it is present only on small scales, is
limited by 
\[
S<S_{LB}=2\pi Er, 
\]
which we will refer to as the \textit{local Bekenstein bound}. It is
interesting to compare this with the entropy of a gas in thermal
equilibrium. One then finds $S_{g}\lesssim Er$ for high temperatures where $%
T\gtrsim 1/r$, and $S_{g}\gtrsim Er$ for low temperatures where $T\lesssim
1/r$. This is quite natural and a consequence of the fact that most of the
entropy in the gas is stored in wavelengths of the order of $1/T$. This
means that the entropy for low temperatures is stored mostly in modes larger
than the Hubble scale and are therefore unaffected by the local Bekenstein
bound $S_{LB}$.

One can also imagine a situation starting out at high temperature, with
entropy stored on small scales. As the universe expands, and the gas is
diluted, the entropy is transferred to larger scales. From the point of view
of the local observer the entropy is no longer stored in local matter,
instead it has receded towards the horizon. The size of the horizon
therefore limits the amount of information on scales larger than the Hubble
scale, or, more precisely, the large scale information that once was
accessible to the observer on small scales. If the horizon is smaller than
its maximal value this is a sign that there is matter on small scales and
the difference limits the entropy (or information) stored in the matter.
This is the role of the D-bound. We conclude, then, that a system with an
entropy in excess of $S_{LB}$ (but necessarily below $S_{H}$) must include
entropy on scales larger than the horizon scale.

\subsection{How does the entropy change?}

\bigskip

While the entropy bounds above are rather well understood, the way entropy
can flow and change involve some rather subtle issues. This is in particular
true when the size of the horizon is slowly changing as in a slowly rolling
inflationary scenario. As a start let us consider a slow roll universe and
compare the horizon size at two different times. We have $r_{1}=r\left(
t_{1}\right) $ and $r_{2}=r\left( t_{2}\right) $ with $r_{2}\gg r_{1}$ for $%
t_{2}\gg t_{1}$. From the previous section it is quite clear that it is
possible to have a much larger entropy within the radius $r_{2}$ at time $%
t_{1}$ than the area of a sphere with radius $r_{2}$ without violating the
Hubble bound. At time $t_{2}$, however, the entropy within $r_{2}$, which
now has become the horizon, is limited by the horizon area. As a
consequence, the entropy that has passed out through the fixed sphere with
radius $r_{2}$ between time $t_{1}$ and $t_{2}$ will exceed the limit set by
the area of the horizon. It is important to realize that there is no
contradiction in this. The crucial issue is how much entropy has passed out
through the apparent horizon which grows from $r_{1}$ to $r_{2}$ during the
time interval, and this will, indeed, be limited by the horizon size at $%
t_{2}$. This is a direct consequence of the D-bound, and is explicit in the
calculation of \cite{Frolov:2002va}.

So far we have discussed entropy in the form of a diluting gas. As the
universe expands this implies a flow of entropy out through the horizon, but
as the gas eventually is completely diluted the flow of entropy taps off.
Whether or not the horizon radius is changing, one will never be able to
violate the Hubble bound or get an entropy flow through an apparent horizon
violating the bound set by the area. But we also need to be able to go back
in time where we will note that the entropy flow increases. The crux of the
matter is, however, that if we go far enough back in time, the energy
density will be comparable with the one of the cosmological constant, and
the space time geometry will be dramatically different. As a consequence, we
can not argue that the total entropy flow will be larger than is allowed for
by the area of the horizon.

A potentially more disturbing situation is obtained if we consider an empty
universe (apart from a possibly changing cosmological constant), which can
be traced arbitrarily far back in time, with entropy generated through the
quantum fluctuations that are of importance for the CMBR. As discussed in
several works, \cite{Brandenberger:1992sr}\cite{Prokopec:1992ia}\cite
{Kruczenski:1994pu}\cite{Polarski:1995jg}\cite{Kiefer:1999sj}\cite
{Gasperini:1992xv}\cite{Gasperini:1993yf}\cite{Gasperini:1993mq}, there is
an entropy production that can be associated with these fluctuations and one
can worry that this will imply an entropy flow out through the horizon that
eventually will exceed the bound set by the horizon. This is the essence of
the argument put forward in \cite{Albrecht:2002xs}.

To understand this better one must have a more detailed understanding of the
cause of the entropy. Entropy is always due to some kind of coarse graining
where information is neglected. In the case of the inflationary quantum
fluctuations we typically imagine that the field starts out in a pure state%
\footnote{%
In standard treatments this is done in the infinite past at energy scales
infinitely higher than the Planck scale. In practice the initial conditions
should be set at a finite scale supplied by quantum gravity. This is,
however, not important for the present discussion.}, with a subsequent
unitary evolution that keeps the state pure for all times. This is true
whether we take the point of view of a local observer or use the global
FRW-coordinates. To find an entropy we obviously must introduce a notion of
coarse graining. Various ways of coarse graining have been proposed, but
they all imply an entropy that grows with the squeezing parameter $r_{k}$.
The squeezing formalism was first applied to cosmology in \cite{Grishchuk:ss}
\cite{Grishchuk:bj}. For large squeezing the entropy in a mode with comoving
momentum $k$ is typically given by, \cite{Brandenberger:1992sr}\cite
{Prokopec:1992ia}\cite{Kruczenski:1994pu}\cite{Polarski:1995jg}\cite
{Kiefer:1999sj}, 
\begin{equation}
S_{k}=2r_{k}.
\end{equation}
The squeezing parameter for a massless scalar in an inflating cosmology
obeys 
\begin{equation}
\sinh ^{2}\left( r_{k}\right) =\frac{1}{4k^{2}\eta ^{2}},
\end{equation}
which for late times (small negative conformal time $\eta $) can be
approximated as 
\begin{equation}
r_{k}\sim \frac{1}{2}\ln \frac{1}{k^{2}\eta ^{2}}=\ln \frac{H}{p}.
\end{equation}
Most of entropy is produced at large scales (when the modes are larger than
the horizon), and, as we will show in a moment, well below the Hubble bound.

This is all in terms of the FRW-coordinates, but let us now take the point
of view of the local observer. In this case the freedom to coarse grain is
more limited. In order to generate entropy we must divide the system into
two subsystems and trace out over one of the subsystems in order to generate
entropy in the other. As an example consider a system with $N$ degrees of
freedom divided into two subsystems with $N_{1}$ and $N_{2}$ degrees of
freedom, respectively, with $N=N_{1}+N_{2}$ and $N_{2}>N_{1}$. If the total
system is in a pure state it is easy to show that the entropy in the larger
subsystem is limited by the number of degrees of freedom in the smaller one,
i.e. $S_{2}<\ln N_{1}$.\footnote{%
A simple proof can be found in \cite{Danielsson:um} in the context of the
black hole information paradox.} Applied to our case, this means that the
entropy flow towards the horizon must be balanced by other matter with a
corresponding ability to carry entropy within the horizon. Since the amount
of such matter is limited by the D-bound, the corresponding entropy flow is
also limited. As a consequence, there can not be an accumulated flow of
entropy out towards the horizon that is larger than the area of the horizon.
This does not mean that inflation can not go on for ever, rather it implies
that the local observer will not be able to do an arbitrary amount of coarse
graining. The analysis of \cite{Frolov:2002va} shows, indeed, how
inflationary quantum fluctuations leave the horizon without actually
affecting the horizon area. It is just the flow of energy, present in the
case of slow roll, that makes the horizon grow. The size of the horizon
should be viewed as the upper limit on the amount of entropy or information.
Whether there is entropy in the fluctuations is not addressed in \cite
{Frolov:2002va} -- this is a matter of a subjective coarse graining. As an
example of how this can happen one can consider the effect discussed in \cite
{GibbHawk:1977}, where it was noted that if the fluctuations are actually
measured this leads to an increase in energy of the detector and the horizon
area shrinks. This implies that entropy is generated from the fluctuation
and stored, not at large scales, but at small.

It is instructive to actually calculate the total amount of entropy in the
fluctuations. We will focus on the entropy on large scales. The total
entropy in a volume with radius $R$ on scales larger than the Hubble scales
is of the order 
\begin{equation}
S\sim R^{3}\int_{1/R}^{H}d^{3}pr_{k}\sim R^{3}\int_{1/R}^{H}dpp^{2}\ln \frac{%
H}{p}\sim \frac{R^{3}}{r^{3}}\ll \frac{R^{3}}{rl_{p}^{2}}\sim S_{H}.
\label{scmb}
\end{equation}
That is, there is only of the order of one degree of freedom per horizon
volume and we are therefore far from any holographic bound. On the other
hand, if we trace the volume $R^{3}$ back in time through the expanding
universe, it will eventually become small enough to be contained within just
one horizon volume. Hence all the entropy in (\ref{scmb}) are carried by
field modes that were once inside of the horizon of a single local observer.
As explained above, this observer will not, however, be able to assign such
a huge entropy to the fluctuations that have left, simply because the
necessary coarse graining requires many Hubble volumes to be achieved. To
summarize: \textit{from a local point of view the production of entropy in
quantum fluctuations is limited by the ability to coarse grain; from a
global point of view entropy is created on scales larger than the Hubble
scale.}

One can also note that the concept of thermalization will be quite different
from the two points of view. The early universe will contain, from the point
of view of the FRW-observer, thermal degrees of freedom as well as non
thermal. It is the non thermal ones (well correlated and information
carrying) that eventually leads to the CMBR-fluctuations. In fact, as
explained in \cite{Kiefer:1999sj}, one should not view the entropy in (\ref
{scmb}) as the actual entropy of the fluctuations. It is more appropriate to
view (\ref{scmb}) as the upper limit on the amount of information that can
be present in the fluctuations. The actual entropy that we should assign to
the fluctuations are much smaller than the value given in (\ref{scmb}) --
the difference is the information contained in the acoustic peaks. From the
point of view of the local observer, both thermal and non-thermal
fluctuations start out on small scales but expand and is eventually
contained in the degrees of freedom of the horizon. From the local observer
point of view this means that they all have thermalized.

\bigskip

\section{The slow roll}

\bigskip

Let us investigate how the above ideas work in more detail in the presence
of a slowly changing Hubble constant in a slow roll universe. The Friedman
equations in the presence of a scalar inflaton is given by 
\begin{equation}
H^{2}=\frac{8\pi l_{p}^{2}}{3}\left( \frac{1}{2}\dot{\phi}^{2}+V\left( \phi
\right) \right) ,  \label{fesl}
\end{equation}
while the equation of motion for the scalar field is given by 
\begin{equation}
\ddot{\phi}+3H\dot{\phi}=-V^{\prime }\left( \phi \right) .
\end{equation}
It is useful to define a slowroll parameter according to 
\begin{equation}
\varepsilon =\frac{1}{16\pi l_{p}^{2}}\left( \frac{V^{\prime }\left( \phi
\right) }{V\left( \phi \right) }\right) ^{2},
\end{equation}
and assume $\varepsilon $ to be small. To lowest, zeroth, order in $%
\varepsilon $ the Hubble constant will be fixed at 
\begin{equation}
H^{2}\sim \frac{8\pi l_{p}^{2}}{3}V\left( \phi \right) ,
\end{equation}
(that is, the potential energy will dominate over the kinetic energy in
contributing to the vacuum energy) while $\phi $ is slowly varying according
to 
\begin{equation}
3H\dot{\phi}\sim -V^{\prime }\left( \phi \right) .
\end{equation}
To first order in $\varepsilon $ there will, however, be a variation in $H$
determined by 
\begin{equation}
\dot{H}=-\varepsilon H^{2},
\end{equation}
which is easily integrated to give 
\begin{equation}
H\left( t\right) =\frac{H_{0}}{1+\varepsilon H_{0}t},
\end{equation}
and 
\begin{equation}
a\left( t\right) =\left( 1+\varepsilon H_{0}t\right) ^{1/\varepsilon }.
\end{equation}
We have chosen the constants of integration such that $H\left( 0\right)
=H_{0}$ and $a\left( 0\right) =1$.

A useful model for a slow roll is to assume a cosmological constant together
with another matter component with an equation of state given by $p=w\rho $,
with $w$ constant. In this setup the first Friedman equation becomes 
\begin{equation}
H^{2}=\frac{8\pi l_{p}^{2}}{3}\left( \rho +\rho _{\Lambda }\right) ,
\end{equation}
where 
\begin{equation}
\rho =\frac{\rho _{0}}{a^{q}},
\end{equation}
with $q=3\left( 1+w\right) $. \ We assume $q>0$; $q=0$ would just give an
additional contribution to the true cosmological constant. The Friedman
equation at $t=0$ now tells us that 
\begin{equation}
H_{0}^{2}=\frac{8\pi l_{p}^{2}}{3}\left( \rho _{0}+\rho _{\Lambda }\right) ,
\end{equation}
while first order in $t$ implies 
\begin{equation}
q\Omega =2\varepsilon ,
\end{equation}
where we have defined 
\begin{equation}
\rho _{0}=\Omega \rho _{c}=\frac{3H_{0}^{2}\Omega }{8\pi l_{p}^{2}}.
\end{equation}
As expected, the energy density in the extra matter is suppressed compared
to the cosmological constant. Let us now compare the area of the final
horizon given by 
\begin{equation}
A_{f}=4\pi \left( \frac{8\pi l_{p}^{2}}{3}\rho _{\Lambda }\right) ^{-2},
\end{equation}
and the (smaller) one at $t=0$ given by

\begin{equation}
A_{0}=4\pi \left( \frac{8\pi l_{p}^{2}}{3}\left( \rho _{0}+\rho _{\Lambda
}\right) \right) ^{-2}.
\end{equation}
A short calculation (assuming $\Omega $ small) reveals that 
\begin{equation}
\Delta A=A_{f}-A_{0}=\frac{2\varepsilon }{q}A_{f}.  \label{da}
\end{equation}
The matter energy contained within the volume is given by 
\begin{equation}
E=\frac{4\pi }{3}\rho _{0}H_{0}^{-3}=\frac{\varepsilon }{q}\frac{r}{l_{p}^{2}%
},
\end{equation}
and we therefore conclude that the entropy contained in the matter is
limited by 
\begin{equation}
S\leq \frac{1}{4}\frac{\Delta A}{l_{p}^{2}}=2\pi EH_{0}^{-1}=2\pi Er.
\end{equation}
This is simply an example of the D-bound at work.

It is instructive to compare with the calculation in \cite{Albrecht:2002xs}
where the horizon was obtained from (\ref{fesl}) either by keeping the
kinetic term (the exact apparent horizon), or dropping the kinetic term (the
slow roll approximation)$.$ This lead to a difference in area given by 
\begin{equation}
\Delta A=A_{f}-A_{0}=\frac{\varepsilon }{3}A_{f},  \label{dak}
\end{equation}
and a corresponding entropy bound. It is easy to see that this is a special
case of the discussion above. The parameter $w$ in the equation of state can
be written 
\begin{equation}
w=\frac{p}{\rho }=\frac{\frac{1}{2}\dot{\phi}^{2}-V}{\frac{1}{2}\dot{\phi}%
^{2}+V}.
\end{equation}
The choice to remove just a kinetic term, as in \cite{Albrecht:2002xs}, from
the Friedman equation corresponds to having matter contributing to $\rho $
with $w=1$, which leads to $q=6$ and therefore agreement between equations (%
\ref{da}) and (\ref{dak}). This, then, indicates that an entropy bound of
the form (\ref{da}) (or (\ref{dak})) should be viewed as a bound on the
entropy contained in the matter \textit{within} the horizon.

\bigskip

\section{Conclusions}

\bigskip

It seems difficult to find clear cut implications of holography in the
context of inflation. As advocated in, e.g., \cite{Easther:1999gk}\cite
{Veneziano:1999ts}\cite{Kaloper:1999tt}, holography does not seem to say
anything more than an appropriate use of the generalized second law of
thermodynamics. Furthermore, the latter, in its physical consequences, is
always a consequence of using standard laws of general relativity and field
theory in a proper way.

To get truly new physics one seems to need more than just a counting of the
degrees of freedom. One such possibility would be to study the concept of
horizon complementarity in a cosmological setting. In the black hole case,
complementarity has been used to resolve the black hole information paradox 
\cite{Susskind:1993if}\cite{Susskind:1993ki}\cite{Susskind:1993mu}\cite
{Susskind:1995qc}. According to complementarity the same physics will look
very different from the point of view of a local observer and from the point
of view of FRW-coordinates. (Actually, one would need to wait until
inflation is over to make real measurements relevant to these coordinates.)
We have already seen how coarse graining and entropy necessarily will be
treated in different ways. While objects passing out through the horizon do
not experience anything dramatic from the point of view of \
FRW-coordinates, the local observer will see how the object experiences
effectively higher and higher temperatures as the horizon is approached, and
eventually observe how the object is boiled to pieces and thermalized. Could
these phenomena, and the way complementarity makes the different pictures
compatible, lead to detectable effects? In \cite{Danielsson:2002td} it is
argued that Poincare recurrences play an important role in resolving
possible paradoxes in a universe where inflation ends. Unfortunately the
analysis also shows that the time scales of inflation make it unlikely that
anything of this will be relevant for observable physics. It would, however,
be interesting to push these ideas further.

Another approach to finding effects of quantum gravity through inflation, is
to use its ability to magnify small scale physics to cosmological scales. In
particular one could hope that subtleties in how the effective initial
conditions for the inflaton are chosen by high energy physics could leave a
detectable imprint on the CMBR. These ideas have been investigated in
several recent works, see \cite{Brandenberger:2002sr}\cite{kinney2003} for
reviews. In order to have any chance of finding something detectable, the
scale of new physics can not be higher than the GUT-scale at $10^{16}$ GeV,
see \cite{Bergstrom:2002yd} for a detailed analysis, which happen to
coincide with the string scale in many realistic heterotic string theories.
One might also speculate that holography would provide a new scale. These
kind of ideas were pursued in \cite{Cohen:1998zx} in an attempt to
understand the cosmological constant in the present universe, but it is easy
to apply their ideas to an inflationary scenario. An obvious way to find a
holographic scale is simply to say that the highest energy scale where field
theory make sense, is limited through $r^{3}\Lambda ^{3}\sim \frac{r^{2}}{%
l_{p}^{2}}$, which leads to an energy scale a bit below Planck scale, that
is, $\Lambda \sim \left( \frac{l_{p}}{r}\right) ^{1/3}\frac{1}{l_{p}}$. In 
\cite{Cohen:1998zx} it was argued that the limit can be improved by
observing that $\Lambda $ should be limited by the highest entropy that you
can have with out forming a black hole. That is, $r^{3}\Lambda ^{3}\sim
\left( \frac{r^{2}}{l_{p}^{2}}\right) ^{3/4}$. This leads to a
characteristic scale given by $\Lambda \sim \left( \frac{l_{p}}{r}\right)
^{1/2}\frac{1}{l_{p}}$. (This is actually the temperature of a gas with an
energy density equal to the cosmological constant.) Using the largest
possible value of $H$, around $10^{14}$ GeV, one finds $\Lambda \sim 10^{16}$
GeV, which is, indeed, a potentially interesting energy scale.\footnote{%
This is the same scale as was argued for in \cite{Albrecht:2002xs} using a
different interepretation than in the present paper of the results in
section 4.} Unfortunately it is difficult to give a firm argument for why
this is a relevant limit. It remains, therefore, a challenge to find
non-trivial limits on inflationary physics imposed by holography.

\bigskip

\section*{Acknowledgments}

The author would like to thank Daniel Domert and Martin Olsson for valuable
discussions. The author is a Royal Swedish Academy of Sciences Research
Fellow supported by a grant from the Knut and Alice Wallenberg Foundation.
The work was also supported by the Swedish Research Council (VR).


\bigskip

\begin{thebibliography}{99}
\bibitem{Brandenberger:2002sr}  R.~H.~Brandenberger, ``Trans-Planckian
physics and inflationary cosmology,'' arXiv:hep-th/0210186.

\bibitem{kinney2003}  W. H. Kinney, ``Cosmology, inflation, and the physics
of nothing,'' Lectures given at the NATO Advanced Study Institute on
Techniques and Concepts of High Energy Physics, St. Croix, USVI (2002)
[arXiv:astro-ph/0301448].

\bibitem{Easther:1999gk}  R.~Easther and D.~A.~Lowe, ``Holography, cosmology
and the second law of thermodynamics,'' Phys.\ Rev.\ Lett.\ \textbf{82},
4967 (1999) [arXiv:hep-th/9902088].

\bibitem{Veneziano:1999ts}  G.~Veneziano, ``Pre-bangian origin of our
entropy and time arrow,'' Phys.\ Lett.\ B \textbf{454}, 22 (1999)
[arXiv:hep-th/9902126].

\bibitem{Kaloper:1999tt}  N.~Kaloper and A.~D.~Linde, ``Cosmology vs.
holography,'' Phys.\ Rev.\ D \textbf{60}, 103509 (1999)
[arXiv:hep-th/9904120].

\bibitem{Bekenstein:ax}  J.~D.~Bekenstein, ``Generalized Second Law Of
Thermodynamics In Black Hole Physics,'' Phys.\ Rev.\ D \textbf{9}, 3292
(1974).

\bibitem{Banks:2000fe}  T.~Banks, ``Cosmological breaking of supersymmetry
or little Lambda goes back to the future. II,'' arXiv:hep-th/0007146.

\bibitem{Banks:2001yp}  T.~Banks and W.~Fischler, ``M-theory observables for
cosmological space-times,'' arXiv:hep-th/0102077.

\bibitem{Hellerman:2001yi}  S.~Hellerman, N.~Kaloper and L.~Susskind,
``String theory and quintessence,'' JHEP \textbf{0106}, 003 (2001)
[arXiv:hep-th/0104180].

\bibitem{Fischler:2001yj}  W.~Fischler, A.~Kashani-Poor, R.~McNees and
S.~Paban, ``The acceleration of the universe, a challenge for string
theory,'' JHEP \textbf{0107}, 003 (2001) [arXiv:hep-th/0104181].

\bibitem{Banks:2001px}  T.~Banks and W.~Fischler, ``An holographic
cosmology,'' arXiv:hep-th/0111142.

\bibitem{Witten:2001kn}  E.~Witten, ``Quantum gravity in de Sitter space,''
arXiv:hep-th/0106109.

\bibitem{Dyson:2002nt}  L.~Dyson, J.~Lindesay and L.~Susskind, ``Is there
really a de Sitter/CFT duality,'' JHEP \textbf{0208}, 045 (2002)
[arXiv:hep-th/0202163].

\bibitem{Dyson:2002pf}  L.~Dyson, M.~Kleban and L.~Susskind, ``Disturbing
implications of a cosmological constant,'' JHEP \textbf{0210}, 011 (2002)
[arXiv:hep-th/0208013].

\bibitem{Myung:2003hb}  Y.~S.~Myung, ``Holographic entropy bounds in the
inflationary universe,'' arXiv:hep-th/0301073.

\bibitem{Hogan:2002xs}  C.~J.~Hogan, ``Holographic discreteness of
inflationary perturbations,'' Phys.\ Rev.\ D \textbf{66}, 023521 (2002)
[arXiv:astro-ph/0201020].

\bibitem{Albrecht:2002xs}  A.~Albrecht, N.~Kaloper and Y.~S.~Song,
``Holographic limitations of the effective field theory of inflation,''
arXiv:hep-th/0211221.

\bibitem{Frolov:2002va}  A.~Frolov and L.~Kofman, ``Inflation and de Sitter
thermodynamics,'' arXiv:hep-th/0212327.

\bibitem{Bekenstein:jp}  J.~D.~Bekenstein, ``A Universal Upper Bound On The
Entropy To Energy Ratio For Bounded Systems,'' Phys.\ Rev.\ D \textbf{23},
287 (1981).

\bibitem{Bekenstein:vm}  J.~D.~Bekenstein, ``Entropy Content And Information
Flow In Systems With Limited Energy,'' Phys.\ Rev.\ D \textbf{30}, 1669
(1984).

\bibitem{Schiffer:et}  M.~Schiffer and J.~D.~Bekenstein, ``Proof Of The
Quantum Bound On Specific Entropy For Free Fields,'' Phys.\ Rev.\ D \textbf{%
39}, 1109 (1989).

\bibitem{Fischler:1998st}  W.~Fischler and L.~Susskind, ``Holography and
cosmology,'' arXiv:hep-th/9806039.

\bibitem{KalyanaRama:1998pk}  S.~Kalyana Rama and T.~Sarkar, ``Holographic
principle during inflation and a lower bound on density fluctuations,''
Phys.\ Lett.\ B \textbf{450} (1999) 55 [arXiv:hep-th/9812043].

\bibitem{Bak:1999hd}  D.~Bak and S.~J.~Rey, ``Cosmic holography,'' Class.\
Quant.\ Grav.\ \textbf{17}, L83 (2000) [arXiv:hep-th/9902173].

\bibitem{Brustein:1999ua}  R.~Brustein, ``The generalized second law of
thermodynamics in cosmology,'' Phys.\ Rev.\ Lett.\ \textbf{84} (2000) 2072
[arXiv:gr-qc/9904061].

\bibitem{Brustein:1999ay}  R.~Brustein, S.~Foffa and R.~Sturani,
``Generalized second law in string cosmology,'' Phys.\ Lett.\ B \textbf{471}
(2000) 352 [arXiv:hep-th/9907032].

\bibitem{Brustein:1999md}  R.~Brustein and G.~Veneziano, ``A Causal Entropy
Bound,'' Phys.\ Rev.\ Lett.\ \textbf{84} (2000) 5695 [arXiv:hep-th/9912055].

\bibitem{Brustein:2000fw}  R.~Brustein, ``Causal boundary entropy from
horizon conformal field theory,'' Phys.\ Rev.\ Lett.\ \textbf{86} (2001) 576
[arXiv:hep-th/0005266].

\bibitem{GibbHawk:1977}  G.~W.~Gibbons, ~S.~W.~Hawking, ~``Cosmological
event horizons, thermodynamics, and particle creation,'' Phys.~Rev. \textbf{%
D\ 15} (1977) 2738.

\bibitem{Bousso:2000md}  R.~Bousso, ``Bekenstein bounds in de Sitter and
flat space,'' JHEP \textbf{0104} (2001) 035 [arXiv:hep-th/0012052].

\bibitem{Grishchuk:ss}  L.~P.~Grishchuk and Y.~V.~Sidorov, ``On The Quantum
State Of Relic Gravitons,'' Class.\ Quant.\ Grav.\ \textbf{6} (1989) L161.

\bibitem{Grishchuk:bj}  L.~P.~Grishchuk and Y.~V.~Sidorov, ``Squeezed
Quantum States Of Relic Gravitons And Primordial Density Fluctuations,''
Phys.\ Rev.\ D \textbf{42} (1990) 3413.

\bibitem{Brandenberger:1992sr}  R.~H.~Brandenberger, V.~Mukhanov and
T.~Prokopec, ``Entropy of a classical stochastic field and cosmological
perturbations,'' Phys.\ Rev.\ Lett.\ \textbf{69}, 3606 (1992)
[arXiv:astro-ph/9206005].

\bibitem{Prokopec:1992ia}  T.~Prokopec, ``Entropy of the squeezed vacuum,''
Class.\ Quant.\ Grav.\ \textbf{10} (1993) 2295.

\bibitem{Kruczenski:1994pu}  M.~Kruczenski, L.~E.~Oxman and M.~Zaldarriaga,
``Large squeezing behavior of cosmological entropy generation,'' Class.\
Quant.\ Grav.\ \textbf{11} (1994) 2317 [arXiv:gr-qc/9403024].

\bibitem{Polarski:1995jg}  D.~Polarski and A.~A.~Starobinsky,
``Semiclassicality and decoherence of cosmological perturbations,'' Class.\
Quant.\ Grav.\ \textbf{13} (1996) 377 [arXiv:gr-qc/9504030].

\bibitem{Kiefer:1999sj}  C.~Kiefer, D.~Polarski and A.~A.~Starobinsky,
``Entropy of gravitons produced in the early universe,'' Phys.\ Rev.\ D 
\textbf{62}, 043518 (2000) [arXiv:gr-qc/9910065].

\bibitem{Gasperini:1992xv}  M.~Gasperini and M.~Giovannini, ``Entropy
production in the cosmological amplification of the vacuum fluctuations,''
Phys.\ Lett.\ B \textbf{301} (1993) 334 [arXiv:gr-qc/9301010].

\bibitem{Gasperini:1993yf}  M.~Gasperini, M.~Giovannini and G.~Veneziano,
``Squeezed thermal vacuum and the maximum scale for inflation,'' Phys.\
Rev.\ D \textbf{48} (1993) 439 [arXiv:gr-qc/9306015].

\bibitem{Gasperini:1993mq}  M.~Gasperini and M.~Giovannini, ``Quantum
squeezing and cosmological entropy production,'' Class.\ Quant.\ Grav.\ 
\textbf{10} (1993) L133 [arXiv:gr-qc/9307024].

\bibitem{Danielsson:um}  U.~H.~Danielsson and M.~Schiffer, ``Quantum
Mechanics, Common Sense And The Black Hole Information Paradox,'' Phys.\
Rev.\ D \textbf{48} (1993) 4779 [arXiv:gr-qc/9305012]. Reprinted in \textit{%
Information theory in physics}, 2000, AAPT, editor W.T. Grandy.

\bibitem{Susskind:1993if}  L.~Susskind, L.~Thorlacius and J.~Uglum, ``The
Stretched horizon and black hole complementarity,'' Phys.\ Rev.\ D \textbf{48%
} (1993) 3743 [arXiv:hep-th/9306069].

\bibitem{Susskind:1993ki}  L.~Susskind, ``String theory and the principles
of black hole complementarity,'' Phys.\ Rev.\ Lett.\ \textbf{71} (1993) 2367
[arXiv:hep-th/9307168].

\bibitem{Susskind:1993mu}  L.~Susskind and L.~Thorlacius, ``Gedanken
experiments involving black holes,'' Phys.\ Rev.\ D \textbf{49} (1994) 966
[arXiv:hep-th/9308100].

\bibitem{Susskind:1995qc}  L.~Susskind and J.~Uglum, ``String Physics and
Black Holes,'' Nucl.\ Phys.\ Proc.\ Suppl.\ \textbf{45BC} (1996) 115
[arXiv:hep-th/9511227].

\bibitem{Danielsson:2002td}  U.~H.~Danielsson, D.~Domert and M.~Olsson,
``Miracles and complementarity in de Sitter space,'' arXiv:hep-th/0210198.

\bibitem{Bergstrom:2002yd}  L.~Bergstr\"{o}m and U.~H.~Danielsson, ``Can MAP
and Planck map Planck physics?,'' JHEP \textbf{0212} (2002) 038
[arXiv:hep-th/0211006].

\bibitem{Cohen:1998zx}  A.~G.~Cohen, D.~B.~Kaplan and A.~E.~Nelson,
``Effective field theory, black holes, and the cosmological constant,''
Phys.\ Rev.\ Lett.\ \textbf{82} (1999) 4971 [arXiv:hep-th/9803132].
\end{thebibliography}
\end{document}